# High-Resolution Neutron Capture and Total Cross-Section Measurements, and the Astrophysical $^{95}$Mo(*n*,γ) Reaction Rate at *s*-process Temperatures

**P. E. Koehler[1] and J. A. Harvey**

*Physics Division, Oak Ridge National Laboratory*
*MS-6356, Oak Ridge, TN, USA*
E-mails: koehlerpe@ornl.gov, harveyja@ornl.gov

**K. H. Guber and D. Wiarda**

*Nuclear Science and Technology Division, Oak Ridge National Laboratory*
*MS-6356, Oak Ridge, TN, USA*
E-mails: guberkh@ornl.gov, wiardada@ornl.gov

Abundances of Mo isotopes predicted by stellar models of the *s* process are, except for $^{95}$Mo, in good agreement with data from single grains of mainstream presolar SiC. Because the meteorite data seemed sound and no reasonable modification to stellar theory resulted in good agreement for $^{95}$Mo, it has been suggested that the recommended neutron capture reaction rate for this nuclide is 30% too low. Therefore, we have made a new determination of the $^{95}$Mo(*n*,γ) reaction rate via high-resolution measurements of the neutron-capture and total cross sections of $^{95}$Mo at the Oak Ridge Electron Linear Accelerator. These data were analyzed with the R-matrix code SAMMY to obtain parameters for resonances up to $E_n$ = 10 keV. Also, a small change to our capture apparatus allowed us to employ a new technique to vastly improve resonance spin and parity assignments. These new resonance parameters, together with our data in the unresolved range, were used to calculate the $^{95}$Mo(*n*,γ) reaction rate at *s*-process temperatures. We compare the currently recommended rate to our new results and discuss their astrophysical impact.

---

[1] Speaker

## 1.    Introduction

Isotopic abundance ratios measured for several trace elements discovered in presolar SiC grains recovered from primitive meteorites represent stringent tests for *s*-process models. These meteorite data greatly expand the number of (effectively) *s*-only isotopes with which models must agree. At the same time, these isotope ratios often are measured with much greater precision than the ratios of *s*-only isotopes for different elements, which previously formed the main constraint on *s*-process models. In a recent paper, Lugaro *et al*. [1] compared abundance ratios predicted by stellar *s*-process models to meteoric data for Sr, Zr, Mo, and Ba. There was good agreement between the data and the models, in general, except for $^{137}$Ba and $^{95}$Mo. The disagreement at $^{137}$Ba also had been noted previously [2] while comparing *s*-process models to isotope ratios measured for aggregates of presolar SiC grains. Because there seemed to be no reasonable way to modify the models to bring $^{137}$Ba in line, it was suggested that the recommended $^{137}$Ba($n,\gamma$) rate was in error. Subsequent measurements [3] however were in agreement with the recommended rate to high accuracy. Hence, $^{137}$Ba remains a puzzle. Lugaro *et al*. also could not find any other way to reconcile the theoretical and measured $^{95}$Mo abundances, so they predicted that the recommended $^{95}$Mo($n,\gamma$) rate was 30% too low at *s*-process temperatures.

The $^{95}$Mo($n,\gamma$) rate used in most *s*-process calculations is the one recommended by Bao *et al*. [4], which is based on the work of Winters and Macklin [5], which, in turn, is based on the measurements of Musgrove *et al*. [6] From their data, Musgrove *et al* calculated a Maxwellian-averaged cross section (MACS) at (only) $kT = 30$ keV of $\langle\sigma\rangle_{30} = 374\pm50$ mb. Winters and Macklin reanalyzed these same data 11 years later and arrived at a MACS which was 30% lower and four times more accurate ($\langle\sigma\rangle_{30} = 292\pm12$). Winters and Macklin offer little explanation for these large changes except that, "…those measurements and attendant data reduction were performed early in the operation of ORELA. The present data reduction relies on the reproducibility of the neutron flux monitor and the near constancy of the ORELA flux shape to reduce the systematic uncertainty to $\pm2$% due to the flux normalization." They further asserted that the cross section ratios for the Mo isotopes should be accurate to better than 1%. Despite these claims, it seemed reasonable to make new measurements given that the original rate of Musgrove *et al* would yield good agreement between the *s*-process models and the meteoric data.

There were several other reasons for making new measurements. First, Musgrove *et al*. [6] did not measure below $E_n = 3$ keV. Although this is acceptable for calculating the MACS at $kT = 30$ keV, it requires a sizeable extrapolation of the data to calculate the MACS at $kT = 8$ keV, where most of the neutron exposure occurs in stellar *s*-process models. Second, there is very limited neutron total cross section data for $^{95}$Mo, but such data can be crucial for obtaining the most accurate MACS. This is because ($n,\gamma$) measurements typically are made with fairly thick samples, so sizeable corrections must be made for resonance self shielding and multiple scattering. Accurate corrections for these effects require accurate resonance neutron widths, which can be determined best from total cross section measurements. Lastly, the resonance analysis of Musgrove *et al.* was made over only a very limited range ($E_n$ = 3–5 keV). To obtain the most accurate MACS, it is desirable to extend the resonance analysis over the widest possible range. For these reasons, we undertook new measurements of the neutron capture and total cross sections for $^{95}$Mo at the Oak Ridge Electron Linear Accelerator (ORELA).

## 2.    The ORELA Experiments

The total neutron cross section for $^{95}$Mo was measured via transmission on flight path 1 at ORELA. The sample was 0.0251 at/b thick and enriched to 96.47% in $^{95}$Mo. A $^6$Li-loaded glass scintillator, 80 m from the neutron-production target, was used to measure transmitted neutrons. The sample was periodically exchanged with an empty sample holder, as well as Bi and $CH_2$ samples, which were used for background determinations.

The $^{95}$Mo($n,\gamma$) cross section was measured at the same time using a new apparatus on flight path 6 in the 40-m station at ORELA. This new apparatus is essentially a clone of the one we have used for several years on flight path 7 [3] to measure neutron capture cross sections to high accuracy. A pair of $C_6D_6$ detectors was used to record γ rays following neutron capture events. The pulse-height-weighting and saturated-resonance techniques were used to convert counts to cross sections. Separate measurements were made with an empty sample holder for background determination. The shape of the neutron flux was determined using a $^6$Li-loaded glass scintillator, placed in the beam approximately 1 m ahead of the sample. The sample was 0.00459 at/b thick by 2.54 cm in diameter and enriched to 96.47% in $^{95}$Mo.

A small change was made in the neutron-capture apparatus to allow pulse-heights (γ-ray energies) to be recorded separately for singles (events in only one detector) and coincidence (simultaneous events in both detectors) events. Although the coincidence rate was low (to avoid systematic uncertainties in the pulse-height weighting technique), this small change was very valuable for determining resonance spin and parity information.

## 3.    Resonance Analysis and Spin-Parity Assignments

The data were analyzed with the R-matrix program SAMMY [7] to determine resonance parameters. Neutron widths for some of the resonances were large enough that parities could be assigned from resonance shapes in the transmission data: *s*-wave resonances have an asymmetric shape due to interference with the relatively large *s*-wave potential scattering, whereas *p*-wave resonances are more symmetric because the *p*-wave potential scattering is much smaller at these energies. Because $^{95}$Mo is near the peak of the *p*- and valley of the *s*-wave neutron strength functions, *p*-wave resonances are visible at very low energies and so six different $J^\pi$ combinations must be considered for each resonance; $1^-$, $2^-$, $2^+$, $3^-$, $3^+$, and $4^-$.

Parities, as well as spins, also could be assigned using the singles and coincidence pulse-height data. Originally, we tried using the technique pioneered by Coceva *et al.* [8], which is based on the expectation that states with higher spin will, on average, emit more γ rays while deexciting to the ground state of $^{96}$Mo. As a result, resonances having higher spin are expected to yield relatively more coincidence counts as well as a softer singles spectrum. Therefore, the ratio of coincidence counts with a low threshold (all coincidences) to singles counts with a high threshold (hard singles) is expected to be correlated with the spin of the resonance. Although this ratio worked fairly well for separating the different spins, it was not useful for separating the two parities (i.e. $2^-$ from $2^+$ and $3^-$ from $3^+$). Examination of the data revealed that there were several other pulse-height ratios that, when used together, allowed many more spins and parities to be assigned. Some of these data are shown in Fig. 1.

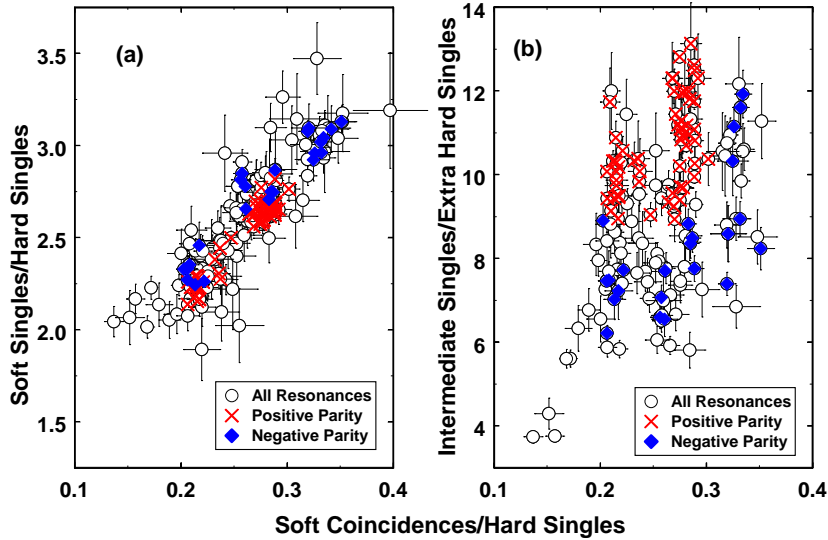

**FIGURE 1.** Ratios of counts in different pulse-height regions in singles and coincidence data taken at ORELA with a $^{95}$Mo sample. Circles represent all resonances below 7 keV. Blue diamonds and red X's depict resonances identified as having negative or positive parity, respectively, by using only our ORELA transmission data. See text for details.

Although the analysis is not yet completed, we already have succeeded in vastly improving the $J^\pi$ assignments for $^{95}$Mo+$n$ resonances. At present, we have been able to make 134 firm $J^\pi$ assignments out of 179 resonances observed below 5.4 keV. In contrast, in Ref. [8], only 10 firm $J^\pi$ assignments were made to 1.14 keV, one of which disagrees with our results. According to compilations [9], there are 13 firm $J^\pi$ assignments below 1.20 keV, all of which agree with our results. Finally, 32 firm $J^\pi$ assignments below 2.05 keV were made in a very recent experiment [10], two of which do not agree with our results. Finally, because part of the technique relies only on singles data, it can be applied to our previous data from ORELA. First tests on data for isotopes of Pt indicate that this new technique, using the singles data alone, will be very useful for assigning both spins and parities.

## 4. The $^{95}$Mo($n,\gamma$) Reaction Rate at *s*-process Temperatures

We used our resonance parameters together with our data in the unresolved region to calculate MACS following standard techniques. The results are compared to previous work in Fig. 2, where it can be seen that our new rates are 20–30% higher than the previously recommended rate at *s*-process temperatures. We estimate overall uncertainties of about 3%. Our new rates should be much more reliable than previous rates for reasons outlined above. Because our new rates are close to what Lugaro *et al*. [1] predicted, there now should be much better agreement between the meteoric data and *s*-process models. However, the recommended rates for [96,97,98]Mo(*n*,γ) also are based on the same reanalysis by Winters and Macklin [5] of the previous data by Musgrove *et al*. [6], and the rates for [96,97]Mo differ by about 10% (in opposite directions) between the two works. Our new data indicate that Winter and Macklin's reanalysis of the [95]Mo data was incorrect. Therefore, new measurements of the reaction rates for [96,97]Mo should be made.

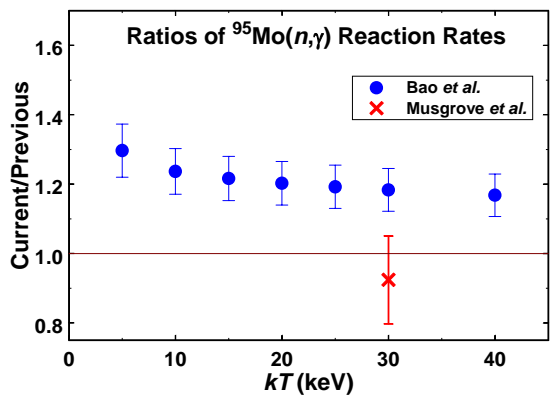

**FIGURE 2.** Ratios of our new [95]Mo(*n*,γ) reaction rates to previous rates. Blue circles and the red X depict ratios of our rates to Bao *et al*. [4] and Musgrove *et al*. [6], respectively.

## 5. Acknowledgements

This work was supported by the U.S. Department of Energy under Contract No. DE-AC05-00OR22725 with UT-Battelle, LLC.

## References

[1] M. Lugaro *et al*., Astrophys. J **593**, 486 (2003).

[2] R. Gallino *et al*., in *Astrophysical Implication of the Laboratory Study of Presolar Materials*, New York: AIP, 1997, p. 115.

[3] P.E. Koehler *et al*., Phys. Rev. C **57**, R1558 (1998).

[4] Z.Y. Bao *et al*., Atomic Data and Nuclear Data Tables **76**, 70 (2000).

[5] R. Winters and R. Macklin, Astrophys. J. **313**, 808 (1987).

[6] A. de L. Musgrove *et al*., Nucl. Phys. **A270**, 108 (1976).

[7] N.M. Larson, Oak Ridge National Laboratory Technical Report ORNL/TM-2000/252 (2000).

[8] C. Coceva *et al*., Nucl. Phys. **A117**, 586 (1968).

[9] S.F. Mughabghab, *Atlas of Neutron Resonances*, Amsterdam: Elsevier, 2006.

[10] S.A. Sheets *et al*., Phys. Rev. C **76**, 064317 (2007).